\documentclass[aps,prb,floats,twocolumn,twoside,floatfix]{revtex4}

\usepackage{epsfig}
\usepackage{amsmath}
\usepackage{amssymb}

\newcommand{\s}{\sigma}
\newcommand{\fann}{f_{\scriptscriptstyle \rm ann}}
\newcommand{\frsb}{f_{\scriptscriptstyle \rm 1RSB}}
\newcommand{\frs}{f_{\scriptscriptstyle \rm RS}}
\newcommand{\sann}{s_{\scriptscriptstyle \rm ann}}

\newcommand{\Tdyn}{T_{\scriptstyle \rm d}}

\newcommand{\TK}{T_{\scriptstyle \rm s}}

\newcommand{\Tc}{T_{\scriptstyle \rm c}}
\newcommand{\Dc}{D_{\scriptstyle \rm c}}
\newcommand{\Dz}{\!{\rm D}z}

\begin{document}

\title{ Inverse Freezing in Mean-Field Structural Glasses }
            
\author{Mauro Sellitto} \affiliation{The Abdus Salam International
  Centre for Theoretical Physics \\ Strada Costiera 11, 34014 Trieste,
  Italy}
 
\begin{abstract}
  A disordered spin model suitable for studying inverse freezing in
  fragile glass-forming systems is introduced.  The model is a
  microscopic realization of the ``random-first order'' scenario in
  which the glass transition can be either continuous or discontinuous
  in thermodynamic sense.  The phase diagram exhibits a first-order
  transition line between two fluid phases terminating at a critical
  point.  When the interacting degrees of freedom are entropically
  favoured an inverse static glass transition and a double inverse
  dynamic freezing appear.
\end{abstract}

\maketitle

Inverse melting and inverse freezing occur when a crystalline or
amorphous solid reversibly transforms into a liquid upon cooling.
This unusual phase behaviour was first predicted by Tammann in 1903,
and is generally considered to be rare because it does involve a
counter-intuitive increase of thermal disorder as the temperature is
lowered~\cite{stillinger}.  An example of biological relevance is
provided by {\em elastin}~\cite{elastin}.  In the past few years
``inverse temperature transitions'' of this type have attracted a
renewed interest as they have been observed in a variety of soft
matter systems including polymers~\cite{Rastogi}, colloids~\cite{Pham}
and micelles~\cite{Malla}.  While the responsible physico-chemical
interactions may depend on the system under study, it has been
recognized that a large enough degeneracy of the degrees of freedom
interacting at low temperature provides a simple mechanism whereby
inverse melting or inverse freezing may generally occur~\cite{ScSh}.

The idea is easily described by considering an ensemble of polymers
that have a low temperature ``folded'' state in which they are
mutually weakly interacting, and a higher temperature ``unfolded''
state which is favored entropically and in which they interact
strongly with each other. As temperature is increased, each polymer
stretches out to reach the other polymers, the resulting entangling
thus may lead to a glass transition.  To obtain a minimal model of
freezing by heating, one can consider~\cite{ScSh} spins taking values
$0, \pm 1$, and a Hamiltonian consisting of a term $\sum_i \s_i^2$
favouring the ``folded'' states $\s_i=0$, and an interaction term
$\sum_{ij} J_{ij} \s_i \s_j$ that is active in the ``unfolded''
states, $\s_i=\pm 1$. The entropic favouring of the latter is enhanced
by making them $r$-fold degenerate.  If the interactions matrix $J$ is
taken from the Gaussian ensemble, one then obtains a reentrant {\em
  spin-glass} phase~\cite{ScSh,CrLe,GS,Chico,BEGSG,Schreiber}.

In this paper, the inverse freezing problem is addressed in the
context of mean-field models of structural glasses by using the above
mechanism of entropy-driven reentrance.  Several reasons make such a
problem interesting.  The ``random first-order'' scenario for the
glass transition~\cite{KiThWo} predicts that upon cooling fragile
glass-forming liquids undergo a purely dynamic arrest before a
thermodynamic singularity occurs at a lower temperature (or higher
density).  The dynamic arrest is the relevant one from an experimental
point of view and, in order to compare observations with theoretical
predictions, one should consider the effect of degeneracy on the
dynamics.  The point is important because a reentrant glass transition
has been recently predicted by mode-coupling theory~\cite{Ken} and
found in colloids~\cite{Pham} and micelles with attractive
interaction~\cite{Malla}.  The second and more general question that
arises concerns the interplay of the two (static and dynamic) glass
transitions and its effect on glassy behaviour~\cite{Leto} when a
reentrance in the phase diagram takes place.
  
In order to answer the above questions, we consider a disordered
system of $N$ spin-1 variables with Hamiltonian
\begin{equation}
  H = - 2\sum_{ij} J_{ij} \s_i \s_j + D \sum_i \s^2_i \,, 
  \qquad \s_i=0,\, \pm 1 \label{H}
\end{equation}
where $J$ is a symmetric random orthogonal matrix (with $J_{ii}=0$),
and $D$ is a crystal field playing a role similar to the chemical
potential: Increasing $D$ will favour $\s_i=0$ states and reduce the
effect of frustration.  The case with binary spin variables ($\s_i =
\pm1$) corresponds to the standard Random Orthogonal Model ({\small
  ROM}) studied by Marinari, Parisi and Ritort ~\cite{MaPaRiII}, which
is known to be glassy at low temperature~\cite{MaPaRiII,Rao,ChDeLe}.
It should be emphasized that quenched disorder is not crucial as
{\small ROM} shares the same basic phenomenology with systems having
deterministic interactions~\cite{MaPaRiII,BoMe}.

The free-energy of the model~(\ref{H}) can be evaluated as in the
standard {\small ROM} by using the replica method and the
identity~\cite{MaPaRiII}: $ \overline{\exp({\rm Tr} \, J A)} = \exp(N
{\rm Tr}\, G(A/N)) $, where $A$ is a symmetric matrix of finite rank,
the overbar is the average over the quenched disorder which is
defined by the Haar measure on the orthogonal group, and
%
\begin{eqnarray}
  G(z) =  \frac{\sqrt{1 + 4 z^2} - 1}{2}  
    -{1\over2} \ln \frac{\sqrt{1 + 4 z^2} + 1}{2} .
\end{eqnarray}
Averaging the replicated partition function gives
\begin{eqnarray}
  \overline{Z^n} &\sim& \int_{-\infty}^{\infty} \prod_{a,b}
  d\Lambda_{ab} dQ_{ab} \exp \left( - \beta f[Q,\Lambda] \right) \,,
\end{eqnarray}
where $a,\, b = 1,\dots,\, n$ are replica indexes, and the Parisi
order parameter $Q_{ab} = \langle \s_a \s_b \rangle$ includes the
diagonal terms, which corresponds to the density of $\pm1$ spins,
$Q_{aa}= \rho$. The free energy $f[Q,\Lambda]$ reads
\begin{equation}
  - \beta f [Q, \Lambda] = {1\over 2}{\rm Tr} G(4 \beta Q) - {\rm Tr}
  (\Lambda Q) + \ln Z_0 \,,
\end{equation}
where $Z_0$ is the single-site partition function
\begin{equation}
  Z_0 = \sum_{\{\s_a\}} \exp\left(\sum_{a,b} \Lambda_{ab} \, \s_a \s_b
  -\beta D \sum_a \s_a^2 \right) .
\end{equation}
In order to proceed one now needs to specify an ansatz for $Q$ and
$\Lambda$, and then consider the zero-replica limit $n\to 0$.

\begin{figure}
\begin{center}
  \includegraphics[width=7.0cm]{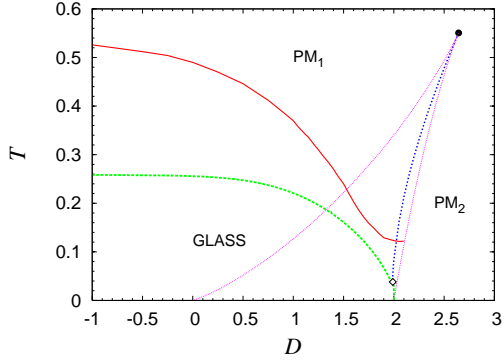}
\end{center}
\caption{(Color online) Phase diagram of {\small ROM} spin-1 in the 
  temperature-crystal field plane.  The continuous (red) line is the
  dynamic freezing, while the dashed (green) line is the static glass
  transition.  The dotted (blue) line is the first-order transition
  between the two paramagnetic (PM$_1$ and PM$_2$) phases.  Spinodals
  are shown as dotted (violet) light lines. The full dot is the
  critical point terminating the coexistence line.  The diamond symbol
  is the tricritical point separating the continuous and discontinuous
  glass transition (at lower temperature).}
\label{fig:TD}
\end{figure}

\smallskip

\noindent
{\bf The fluid-fluid transition.}  The simplest case is the replica
symmetric ansatz: $ Q_{ab}=(\rho-q)\delta_{ab}+q$, $\Lambda_{ab}
=(\mu-\lambda)\delta_{ab}+\lambda$, where $\delta$ is the Kronecker
symbol.  In this approximation the free energy reads:
\begin{eqnarray}
  \beta \frs & =& -\frac{1}{2}
  G(4\beta(\rho-q)) - 2\beta q G'(4\beta(\rho-q)) - \lambda q  + \mu \rho
   \nonumber \\ & & - \int \Dz \ln \left(1+ 2 {\rm
  e}^{\mu-\lambda-\beta D} \cosh ( z\sqrt{2 \lambda} ) \right),
\end{eqnarray}
where $\Dz \equiv dz {\rm e}^{-z^2/2}/\sqrt{2\pi}$ and $q$, $\rho$,
$\lambda$ and $\mu$, are self-consistently determined by the
saddle-point equations.  At sufficiently large temperature/crystal
field, $q=\lambda=0$, and one recovers the annealed free energy
\begin{eqnarray}
  \beta \fann &=&
  -{1 \over 2} G(4\beta\rho) + \beta D \rho -s_0(\rho) \,,
\end{eqnarray}
where $s_0(\rho) = -(1-\rho) \log(1-\rho) - \rho \log \rho +\rho \log
2$ is the entropy of a noninteracting spin-1 system, and the density
$\rho=\rho(\beta,D)$ satisfies the implicit equation
\begin{equation}\label{rho_implicit}
  \beta D = \log \frac{2 (1-\rho)}{\rho} 
  + 2\beta G'(4 \beta \rho) \,.
\end{equation}
Equation~(\ref{rho_implicit}) exhibits multiple solutions. There is an
unstable phase (with negative susceptibility) and two paramagnetic
fluid phases (${\rm PM}_1$ and ${\rm PM}_2$) between which a
first-order transition occurs. The latter terminates at a critical
point located at $\Dc \simeq 2.644$, $\Tc \simeq 0.5506$, see
Fig.~\ref{fig:TD}.
The critical point does not appear in spin-1 {\small
  REM}~(Ref.~\onlinecite{Mo}), showing that the two models are not
equivalent in the large-$D$ regime.
Along the first-order transition line two distinct phases having equal
free energy coexist, and in both it is possible to go continuously
around the critical point from one coexisting phase to the other by
appropriately varying $D$ and $T$.  This first-order transition is a
general feature of spin-1 and lattice-gas systems with disordered
(Gaussian or orthogonal) interactions, though it is sometimes missed.
For $D \ge 2$ the density of $\pm 1$ spins decreases upon cooling.
This behaviour is somehow unexpected if compared to what happens in
the Ghatak-Sherrington model, i.e.  Eq.~(\ref{H}) with Gaussian
disorder.  In the latter, the density increases upon cooling and the
critical point is absent.

\begin{figure}
\begin{center}
  \includegraphics[width=7.0cm]{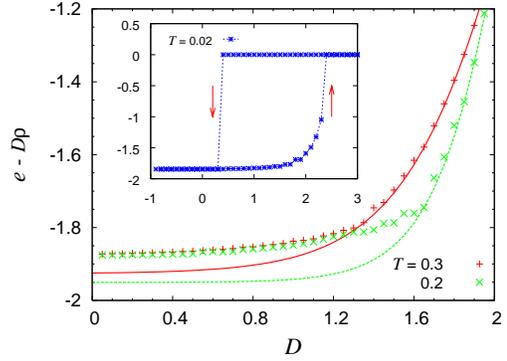}
\end{center}
\caption{(Color online) Energy density $e$ vs crystal field $D$ at fixed
  temperature $T$, in a slow crunching MC experiment. The lines are the
  analytic results of the annealed approximation.}
\label{fig:e_D}
\end{figure}

\smallskip

\begin{figure}
\begin{center}
  \includegraphics[width=7.0cm]{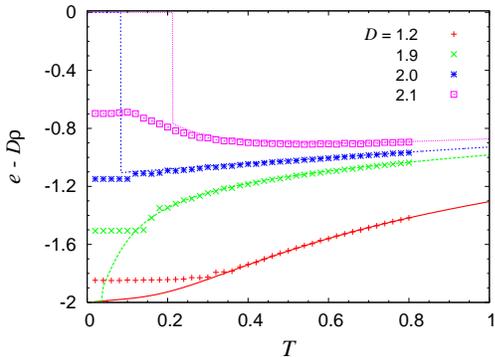}
\end{center}
\caption{(Color online) Energy density $e$ vs. temperature $T$ at fixed  
  crystal field $D$, in a slow cooling MC experiment. The lines are the
  analytic results of the annealed approximation.  }
\label{fig:e_T}
\end{figure}

\noindent
{\bf The static and dynamic glass transition.}
At low temperature/crystal field, the annealed entropy,
\begin{eqnarray}
  \sann &=& 
  s_0(\rho) + {1 \over 2} G(4 \beta \rho) -2 \beta \rho G'(4 \beta \rho)
\,,
\end{eqnarray}
becomes negative, suggesting that replica symmetry has to be broken.
Within the one-step replica-symmetry-breaking ansatz~\cite{MPV}, $Q$
and $\Lambda$ are block diagonal matrices, where the blocks have size
$m \times m$.  Inside the blocks $Q_{ab}=(\rho-q)\delta_{ab}+q$,
$\Lambda_{ab}=(\mu-\lambda)\delta_{ab}+\lambda$. Then the free energy
becomes
\begin{eqnarray}
  \beta \frsb &=& \frac{1-m}{2m}G(4 \beta(\rho-q))
   + \lambda q (m-1)   \nonumber \\
  && -\frac{1}{2m}G(4\beta(\rho-q+qm)) + \ \mu \rho  \\
  \nonumber &&  -\frac{1}{m} \ln \int \Dz \left(
  1+2{\rm e}^{\mu-\lambda-\beta D } \cosh(z\sqrt{2\lambda}) \right)^m \,.
\end{eqnarray}
Expanding near $m=1$ gives $ \beta \frsb \simeq \beta \fann -(m-1) V
$, that allows to locate the static and dynamic transition through the
effective potential $V = - \beta \left.{\partial \frsb \over\partial
    m}\right|_{m=1}$ (Refs.~\onlinecite{KiThWo} and \onlinecite{ChDeLe}).
The two glass transition lines $\TK(D)$ and $\Tdyn(D)$ are shown in
Fig.~\ref{fig:TD}.  In the Ising spin limit, $D \to -\infty$, we get
$\TK \simeq 0.26$ and $\Tdyn \simeq 0.535$, consistently with
Refs.~\onlinecite{MaPaRiII} and \onlinecite{ChDeLe}.  Several
interesting features can be observed.
(i) The temperature at which the annealed entropy vanishes is very
close to $\TK(D)$ for any value of $D$, meaning that the glassy phase
of our model is similar to that of spin-1 {\small REM}
(Ref.~\onlinecite{Mo}).
(ii) There is a tricritical point at $T^* \simeq 0.036$, $D^* \simeq
2.0$ below which the nature of the glass transition changes from
second to first order in the thermodynamic sense.  No appreciable
irreversibility effects are observed across the second-order
glass-transition line when $D$ loops start from the PM$_1$ phase,
Fig.~\ref{fig:e_D} (main frame), whereas there is latent heat and
hysteresis across the first-order glass-transition line, see
Fig.~\ref{fig:e_D} (inset).
(iii) The dynamic freezing line penetrates the ${\rm PM}_2$ phase up to
the spinodal line (at $D \simeq 2.1$), see Fig.~\ref{fig:TD}.  For $D
< 2.1$ the system is dynamically unable to reach equilibrium at low
temperature even in the case in which the ground is trivial, that is
for $2< D < 2.1$, see Fig.~\ref{fig:e_T}.
(iv) Crunching the system from high to low $D$ at fixed temperature,
$\Tdyn > T > \TK$, leads to a purely dynamic freezing with no
underlying entropy crises.
Monte Carlo (MC) results shown in Figs.~\ref{fig:e_D} and ~\ref{fig:e_T}
are for a system of size $N=512$ and a very slow annealing rate
($10^7$ MC sweeps per unit variation of $D$ and $T$, respectively).

\begin{figure}
\begin{center}
  \includegraphics[width=7.0cm]{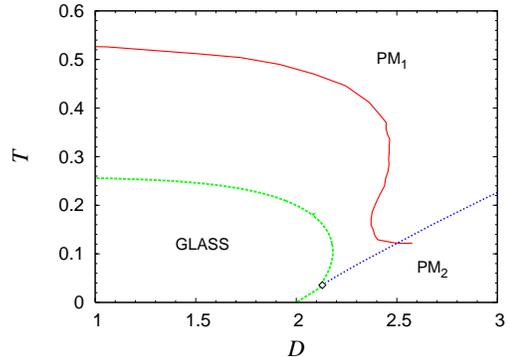}
\end{center}
\caption{(Color online) Phase diagram of {\small ROM} spin-1 for 
  degeneracy ratio $r=50$.  Spinodal lines are not shown here.}
\label{fig:TD_r50}
\end{figure}

\smallskip

\noindent
{\bf Inverse temperature glass transitions.} To take into account
inverse freezing phenomena, the interacting states, $\s_i=\pm 1$, are
now given an entropic advantage by a degeneracy ratio $r>1$ with
respect to the noninteracting states, $\s_i=0$. The opposite case in
which $r<1$ will not be discussed here: That would lead only to a
reentrance of PM$_1$ within the PM$_2$ fluid phase, which, for our
purpose, is less interesting.  One can easily see that including
degeneracy in the system corresponds to changing the crystal field in
Eqs. (6--10) as follows: $D \rightarrow D - T \log r$
(Ref.~\onlinecite{ScSh}).  In fact, the dependence of the effective
potential on the degeneracy ratio enters only through the density
variable, $\rho$, which satisfies Eq.~(\ref{rho_implicit}).  Having
the meaning of the different phases already clarified, we need only to
investigate how the phase boundaries in Fig.~\ref{fig:TD} are modified
accordingly.
The analysis of the effective potential shows that for small $r$ the
reentrancy effect is weak, while for large enough values of $r$ a
rather interesting reentrant behaviour appears. We discuss the results
for the case $r=50$, which corresponds to the phase diagram presented
in Fig.~\ref{fig:TD_r50}. First of all, we see that the two
glass-transition lines never cross each other and that upon heating
the inverse static glass transition occurring in the range $2<D<2.18$
is anticipated by inverse freezing.  This is due to the penetration of
the dynamical arrest line in the PM$_2$ fluid phase.
At intermediate crystal field, $2.18<D<2.37$, there is inverse
freezing without a static glass transition. This prevents the
application of the Kauzmann paradox, since in this range of crystal
field the equilibrium phase at low temperature is the PM$_2$ fluid.
Finally, at larger crystal field, $2.37<D<2.58$, there is a {\em
  double} inverse dynamical freezing (again with no underlying static
glass transition) that goes through the PM$_1$ and PM$_2$ fluid phases
(i.e., on cooling one would observe the sequence of transitions
PM$_1$-G-PM$_1$-G-PM$_2$). In this region of the phase diagram the
packing density of $\pm 1$ spins decreases with the temperature, i.e.,
the glass state at lower temperature is less dense than the one at
higher temperature.  This behaviour is reminiscent of that predicted
by mode-coupling theory~\cite{Ken} and observed in experiments on
attractive colloids~\cite{Pham} and micelles~\cite{Malla}.

\smallskip

\noindent
{\bf Conclusions.}  To summarize, we introduced a generalisation of
{\small ROM} (Ref.~\onlinecite{MaPaRiII}) allowing investigation of
inverse freezing phenomena in fragile glass-forming liquids through a
mechanism of entropy-driven phase reentrance~\cite{ScSh}.  The model
is a microscopic realization of the ``random first-order'' scenario
for the the structural glass transition~\cite{KiThWo}.  However, in
our case the glass transition can be either continuous or
discontinuous in thermodynamic sense.  This is due to the presence of
a first-order transition between two fluid phases.
Similar results are obtained by using purely biquadratic interactions,
$J_{ij} \s_i^2 \s_j^2$, that would correspond to a lattice-gas {\small
  ROM} ($\s_i^2 \to n_i=0,1$).
Notice that our results have no counterpart in the quantum version of
{\small ROM} studied in Ref.~\onlinecite{Ri}.  Rather, they have some
resemblance with those obtained in Refs.~\onlinecite{Go} and
\onlinecite{CuGrSa}, reproducing a number of experimental observations
on the dipolar spin-glass LiHo$_x$Y$_{1-x}$F$_4$ in external
field~\cite{dipolar}. In fact, the present model can be considered as
the insulating limit of a generalisation of the itinerant electron
model (see Ref.~\onlinecite{FeOp}) with random orthogonal
interactions.

At large enough degeneracy an inverse static glass transition and a
double inverse dynamic freezing occur. The latter reproduces
qualitatively some features observed in recent experiments on
colloidal and copolymer-micellar systems with short-range
attraction~\cite{Pham,Malla}. Nevertheless, the possibility of
describing the glass-glass transition~\cite{Ken,Cates,Ema_Zac} in the
present setting remains unclear (see, however,
Refs.~\onlinecite{Antonio} and \onlinecite{CrLe2}) and deserves
further investigation.
Let us finally mention that this work can be extended by considering
both bilinear and biquadratic interactions, similarly to what has been
done in Ref.~\onlinecite{BEGSG}. One can also include a three-body
interaction term to mimick microemulsion (see, e.g.,
Refs.~\onlinecite{GoSc}).  Depending on the relative strength of these
interactions and the nature of the quenched (Gaussian versus
orthogonal) disorder an even richer variety of phases is expected.
That could lead to a better understanding of the glassy behaviour of
complex liquids and soft matter systems.

\smallskip

It is a pleasure to thank L. Cugliandolo, D. Dean, and especially J.
Kurchan for discussions and suggestions.  Support of the {\small
  EVERGROW} project is acknowledged.

\end{document}